# Photometric investigations of distant comets C/2002 VQ94 (LINEAR) and 29P/Schwassmann–Wachmann-1

**A. V. Ivanova[a], P. P. Korsun[a], and V. L. Afanasiev[b]**

[a] Main Astronomical Observatory of the National Academy of Sciences of Ukraine
[b] Special Astrophysical Observatory, Russian Academy of Sciences, Russia

**Abstract**—We present an analysis of the results of photometric investigations of two distant comets, C/2002 VQ94 (LINEAR) and 29P/Schwassmann–Wachmann-1, obtained with the 6-m telescope of the Special Astrophysical Observatory of the Russian Academy of Sciences. The comets under study demonstrate sufficient activity out of the zone of water ice sublimation (at heliocentric distances longer than 5 AU). In the spectra of the investigated comets, we found the $CO^+$ and $N_2^+$ emission. The presence of this emission may say that the comets were formed in the outer parts of the Solar System, in a protoplanetary cloud at a temperature ≤25 K. We found that the photometric maximum of the ionosphere (in the $CO^+$ filter) of the comet C/2002 VQ94 (LINEAR) is shifted relative to the photometric center of the dust coma by 1.4″ (7.44 × 10$^3$ km) in the direction deflected by 63° from the direction to the Sun. Using special filters to process the images, we picked out active structures (jets) in the dust coma of the 29P/Schwassmann–Wachmann-1 comet.

## INTRODUCTION

A comet as a phenomenon usually forms when a comet nucleus approaches the Sun at a sufficient enough distance to start the process of sublimation (<3 AU) of water ice, this being the main volatile component of the nuclei. Further, the comet head forms from neutral water vapor and admixtures. The sublimate stream from the comet surface forms a dust tail carrying refractory particles.

Nonetheless, it is known from observations that some comet nuclei show activity at heliocentric distances far exceeding the boundary distance (Sekania et al., 1992; Rauer et al., 1997) where the water ice sublimation happens. Various mechanisms to explain the nucleus activity at long heliocentric distances were proposed. Many authors explain the comet activity at long heliocentric distances by the sublimation of the most volatile admixtures, namely, CO and/or $CO_2$ ice (Prialnik, Bar-Num, 1992), or by the combination of exothermal processes in the nucleus such as, for example, the polymerization of HCN and crystallization of the amorphous water ice (Gronkowski, Smela, 1998; Capria, 2002; Prialnik, 2002).

The subjects of our study are comets with perihelion distances longer than 5 AU (they never enter the water ice sublimation region). Their activity is connected with the formation of long tails. As a rule, those tails have no internal structure, their width is approximately the same along the tail, they often are strongly curved, and the position angle of the prolonged radius vector may be up to 60°.

During observations of these comets, it was noted that two comets have specific differences from the other comets of our sample. Processing observations, we found that comets C/2002 VQ94 (LINEAR) (further VQ94) and 29P/Schwassmann–Wachmann-1 (further SW1) have similar morphology and in the spectra of those comets' $CO^+$, $N_2^+$, CN, and $C_3$ bands, not including the continuum. We should note that for the first time the $CO^+$ lines were identified during observations carried out in 1978 and 1979 by Cocran et al. (1980). Further, Jokers et al. (1992) obtained images of the SW1 comet in the $CO^+$ line and investigated the morphology of its ionic component.

As for the morphology of the comets, both of them show such activity as an asymmetric coma and features such as jets, in spite of the fact that they do not enter the water ice sublimation region.

Both of the comets have different dynamic behavior, but the same physical properties. The periodical comet SW1 was discovered in the vicinity of Jupiter ($q = 5.47$ AU and $e = 0.15$) in November 1927. Over the course of time, its orbit underwent evolutionary changes because of Jupiter's attraction influence and became almost circular with parameters of $q = 5.72$ AU and $e = 0.044$. Since the comet was discovered, its orbital period has also changed; it decreased from 16 years down to 14.9 years. Today, the comet is related to the Centaur class.

The comet C/2002 VQ94 (LINEAR) was discovered as a new object of the 19th magnitude in the frame of the investigation project LINEAR in November 2002 when it was at a heliocentric distance of 10 AU and a geocentric distance of 9.16 AU (Marsden, 2002). This object initially was observed by Tegler in 2003 as an asteroid. For the first time, the comet activity of the C/2002 VQ94 (LINEAR) asteroid was registered in August 2003 when it was at a heliocentric distance of 8.9 AU. In images obtained with the 2.2-m telescope of the University of Hawaii (Green, 2003), it was found that the asteroid VQ94 has a notable extensive coma of 10″ with a fan-like structure. In 2003, the object C/2002 VQ94 (LINEAR) was assigned to the comet class (Parker, 2003). The more extensive asymmetric coma of this comet was observed later (in March 2006) in broadband filters $V$ and $R$ at a heliocentric distance of 6.8 AU (Korsun et al., 2006) and medium-band filters in April 2007 at a heliocentric distance of 7.33 AU (Korsun et al., 2008).

Both of the comets have nuclei with dimensions larger than for the typical short-period comets. Using the results of investigations of the SW1 comet with the method of the optical photometry (Meech et al., 1993) and the results of the analysis of the thermal emission (Cruikshank, Brown, 1983; Stanberry et al., 2004), the comet's radius was estimated to be 15–44 km. The effective radius of the VQ94 comet, calculated by Jewitt (2005), is 40.7 km on the condition that its albedo is 0.04. This estimate was obtained when the VQ94 object was not yet assigned as a comet.

The comet SW1 demonstrates occasional outburst activity during many observational periods (Roemer, 1958; Whipple, 1980). In turn, Jewitt (Jewitt, 1990) noted that the coma of SW1 never disappeared completely, in spite of the different level of activity during the entire observational time. According to the data of SW1 monitoring (Trigo–Rodriguez, 2008) from 2002 to 2007, the authors have found 28 outbursts. The average number of outbursts is 7.3 per year. In spite of good and stable observations, it appears impossible to establish an exact periodicity of outbursts. It proves that the activity character of this comet is unpredictable. The origin and orbital evolution of the comet are also not clear yet.

OBSERVATIONS AND DATA PROCESSING

Observations of the SW1 and VQ94 comets were carried out in the frame of the program of spectral and photometric investigations of comets demonstrating high activity at long heliocentric distances. The observation were carried out with the 6-m telescope BTA (SAO RAS, Russia) using the SCORPIO focal reducer mounted on a direct focus (Afanasiev, Moiseev, 2005). The SCORPIO focal reducer was used in photometric and spectral (with a long slit) modes. As a radiation detector, we used the CCD camera EEV-42-40 with a size of 2048 × 2048 pixels. The camera's field of view is 6.1′ × 6.1′ with the scale of the image of 0.18″ per pixel.

The comet SW1 was observed with the broadband $B$, $V$ and $R$ filters in December 2006 during a high activity period of the comet and in the comet's CO$^+$ ($\lambda$ = 4280 Å, FWHM = 40 Å) and the medium-band SED537 ($\lambda$ = 5309 Å, FWHM = 169 Å) filters in November 2007 when the comet demonstrated higher activity as well. The heliocentric and geocentric distances in 2006 were 5.87 and 4.95 AU, respectively, and in 2007, 5.97 and 5.19 AU.

The comet VQ94 was observed with the broadband filter $R$ and the medium-band filters SED415 ($\lambda$ = 4203 Å, FWHM = 212 Å) and SED537 ($\lambda$ = 5309 Å, FWHM = 169 Å) in April 2007. The heliocentric and geocentric distances of the comet in 2007 were 6.8 and 6.6 AU, respectively. More detailed information about the observations is given in the table.

During the entire observational period, due to the SCORPIO device, we could trace the comet motion relative to motionless stars. To compensate for the probable visibility changes which might appear during the observational period with different filters, the sequence of the filters was SED537, SED415, SED537, etc. (see the table).

The obtained data were reduced using programs written in IDL. The images were cleaned from cosmic particle marks. When we reduced the images, we also took into account the bias. Then, the images underwent flat field correction. To compensate for the photometric differences of the detector pixels, we used images of the morning sky. To increase the S/N ratio of the observed date, the photometric images were binned as 2 × 2 (the charges of adjacent cells are summed into one superpixel). In addition, the entire series of homogeneous data were summed for further use.

For the photometric calibration of our data, we observed the spectro-photometric standard stars BD + 75°325, BD + 33°325, and G193-74 (Oke, 1990). The value of the atmospheric transparency in the SAO RAS region we took from a paper by Kartasheva and Chunakova (1978).

COLOR INDEX AND DUST PRODUCTIVITY

As mentioned above, the observations of SW1 and VQ94 comets were carried out using broadband ($V$, $B$, $R$), medium-band (SED415, SED537), and comet (CO$^+$) filters. Using the results of the observations of December 2006 for the SW1 comet, we obtained a color index for the aperture radius of 5.4″.

The values we obtained are $B–V = 0.64 \pm 0.09$ and $V–R = 0.54 \pm 0.08$. If one compares the color indices we obtained with the solar color indices ($B–V = 0.62$ and $V–R = 0.36$ (Drilling, Landolt, 2000)), then the obtained results show that the color of the comet near-nucleus region is closer to red.

According to the results obtained for SW1 in 1976 and 1977 (Kiselev, Chernova, 1979), the color index $B–V$ varied in a range from 0.68 to 1.06. The author

Observational journal

| Moment, UT | $r$, a. e. | $\Delta$, a. e. | Exposure time, s | $P^*$, deg | Filter | Comet |
|---|---|---|---|---|---|---|
| December, 15.911, 2006 | 5.865 | 4.947 | 300 | 102.4 | $B$ | SW1 |
| December, 15.915, 2006 | 5.865 | 4.947 | 300 | 102.4 | $B$ | SW1 |
| December, 15.919, 2006 | 5.865 | 4.947 | 200 | 102.4 | $V$ | SW1 |
| December, 15.922, 2006 | 5.865 | 4.947 | 200 | 102.4 | $V$ | SW1 |
| December, 15.924, 2006 | 5.865 | 4.947 | 100 | 102.4 | $R$ | SW1 |
| December, 15.926, 2006 | 5.865 | 4.947 | 100 | 102.4 | $R$ | SW1 |
| December, 15.996, 2006 | 5.865 | 4.947 | 300 | 102.4 | $B$ | SW1 |
| December, 15.999, 2006 | 5.865 | 4.947 | 200 | 102.4 | $V$ | SW1 |
| December, 16.002, 2006 | 5.865 | 4.947 | 200 | 102.4 | $V$ | SW1 |
| December, 16.005, 2006 | 5.865 | 4.947 | 100 | 102.4 | $R$ | SW1 |
| December, 16.006, 2006 | 5.868 | 4.947 | 100 | 102.4 | $R$ | SW1 |
| April, 9.997, 2007 | 6.834 | 6.685 | 40 | 202.5 | $R$ | VQ94 |
| April, 17.025, 2007 | 6.834 | 6.685 | 120 | 192.3 | SED415 | VQ94 |
| April, 17.027, 2007 | 6.834 | 6.685 | 120 | 192.3 | SED537 | VQ94 |
| April, 17.031, 2007 | 6.834 | 6.685 | 120 | 192.3 | SED415 | VQ94 |
| April, 17.033, 2007 | 6.834 | 6.685 | 120 | 192.3 | SED537 | VQ94 |
| April, 17.035, 2007 | 6.834 | 6.685 | 120 | 192.3 | SED415 | VQ94 |
| April, 17.037, 2007 | 6.834 | 6.685 | 120 | 192.3 | SED537 | VQ94 |
| April, 17.040, 2007 | 6.834 | 6.685 | 120 | 192.3 | SED415 | VQ94 |
| April, 17.042, 2007 | 6.834 | 6.685 | 120 | 192.3 | SED537 | VQ94 |
| April, 17.045, 2007 | 6.834 | 6.685 | 120 | 192.3 | SED415 | VQ94 |
| April, 17.047, 2007 | 6.834 | 6.685 | 120 | 192.3 | SED537 | VQ94 |
| April, 17.049, 2007 | 6.834 | 6.685 | 120 | 192.3 | SED415 | VQ94 |
| April, 17.051, 2007 | 6.834 | 6.685 | 120 | 192.3 | SED537 | VQ94 |
| November, 17.9946, 2007 | 5.967 | 5.197 | 10 | 264.1 | $V$ | SW1 |
| November, 17.9965, 2007 | 5.967 | 5.197 | 10 | 264.1 | $V$ | SW1 |
| November, 17.9979, 2007 | 5.967 | 5.197 | 10 | 264.1 | $V$ | SW1 |
| November, 18.0108, 2007 | 5.967 | 5.197 | 10 | 264.1 | $V$ | SW1 |
| November, 18.0235, 2007 | 5.967 | 5.197 | 10 | 264.1 | $V$ | SW1 |
| November, 18.0246, 2007 | 5.967 | 5.197 | 10 | 264.1 | $V$ | SW1 |
| November, 18.0374, 2007 | 5.967 | 5.197 | 10 | 264.1 | $V$ | SW1 |
| November, 18.0384, 2007 | 5.967 | 5.197 | 10 | 264.1 | $V$ | SW1 |
| November, 18.0662, 2007 | 5.967 | 5.197 | 300 | 264.1 | $CO^+$ | SW1 |
| November, 18.0635, 2007 | 5.967 | 5.197 | 150 | 264.1 | SED537 | SW1 |
| November, 18.0667, 2007 | 5.967 | 5.197 | 300 | 264.1 | $CO^+$ | SW1 |
| November, 18.0860, 2007 | 5.967 | 5.197 | 300 | 264.1 | $CO^+$ | SW1 |
| November, 18.0929, 2007 | 5.967 | 5.197 | 300 | 264.1 | $CO^+$ | SW1 |
| November, 18.0968, 2007 | 5.967 | 5.197 | 150 | 264.1 | SED537 | SW1 |
| November, 18.0991, 2007 | 5.967 | 5.197 | 300 | 264.1 | $CO^+$ | SW1 |
| November, 18.1028, 2007 | 5.967 | 5.197 | 300 | 264.1 | $CO^+$ | SW1 |
| November, 18.1065, 2007 | 5.967 | 5.197 | 300 | 264.1 | $CO^+$ | SW1 |
| November, 18.1106, 2007 | 5.967 | 5.197 | 150 | 264.1 | SED537 | SW1 |
| November, 18.1125, 2007 | 5.967 | 5.197 | 150 | 264.1 | SED537 | SW1 |

* Position angle of the prolonged radius vector given in degrees.

supposed that such variations might be due to a comet outburst.

Color index estimates for SW1 made by another author later gave values of $B-V = 0.8$ (Hartmann et al., 1982) and $V-R = 0.502$ (Meech et al., 1993).

Some authors (Jewitt, 1999; Luu, Jewitt, 1996; Tegler, Romanishin, 2007) suppose that the index color value variations may be explained by several reasons. According to them, it may be a consequence of evolutionary processes or variation connected directly with active processes on the comet nucleus surface.

Estimates of the VQ94 object when it was observed as an asteroid were $B-V = 0.92$, $V-R = 0.47$, and $B-R = 1.39$ (Tegler et al., 2003); our estimate, obtained when the object began to demonstrate comet activity in December 2006, was $V-R = 0.41 \pm 0.07$. Our results show that the value of the object's color index is close to neutral.

To calculate the dust productivity rate of the comets, we used a relation presented in a paper by Meech and Weawer (1996):

$$Q_{dust} = \frac{Af\rho(4/3\pi a^3 \sigma v_{dust})}{\pi a^2 p},$$

where $a$ is the mean radius of a dust particle, $\sigma$ is the density of a dust particle, $v$ is the velocity of the dust particle ejection, and $p$ is the dust particle albedo. In the relation, the multiplication $Af\rho$ is also used, where $A$ is the albedo, $f = \frac{sN_d(\rho)}{\pi\rho^2}$ is the filling factor representing the degree of the diaphragm filling by the comet coma dust particles in projection to the celestial sphere, and $\rho$ is the aperture radius. Since $f \propto N_d(\rho)/\rho^2$ and $N_d(\rho) \propto \rho$, then $Af\rho$ does not depend on the diaphragm size. The relation for $Af\rho$ is written as follows (A'Hern, 1984):

$$Af\rho = \frac{(2r\Delta)^2}{\rho}\frac{F_{com}}{F_{Sun}}. \quad (1)$$

Here, $r$ (AU) is the heliocentric distance of a comet, $\Delta$ (cm) is the geocentric distance of a comet, $\rho$ is the aperture radius of a comet, $F_{com}$ (cm erg$^{-2}$ s$^{-1}$ Å$^{-1}$) is the comet flux in continuum, and $F_{Sun}$ (cm erg$^{-2}$ s$^{-1}$ Å$^{-1}$) is the solar flux at a distance of 1 AU. The value of the intensity of the solar flux has been taken from a paper by Neckel and Labs (1984).

Relation (1) is widely used to determine the relative measure of the dust productivity in different comets in the condition of the isotropic matter outflow to the comet atmosphere. For our objects, the matter outflow is mostly concentrated in jets, hence the value of $Af\rho$ can vary depending on the size of the diaphragm we choose for our calculations. To estimate the measure of the dust productivity of distant comets using the matching method, we determined the value of the aperture radius, in limits of which the $Af\rho$ parameter does not vary or varies slightly. To determine the radius we need, we calculated the $Af\rho$ parameter for different values of the aperture radius (1.8″, 2.52″, 3.24″, 3.96″, 4.86″, 5.4″, 6.12″, 6.84″, 7.56″, 8.28″, and 9.0″) and looked how its value changes depending on the distance from the nucleus center. We obtained that for the SW1 comet, the value of $Af\rho$ changes slightly (less than 11 %) when the radius changes in limits from 3.96″ to 8.28″. To calculate the dust productivity of the SW1 comet, we took the mean value of the radius as 5.4″. For the VQ94 comet, we performed the same calculations and found that the $Af\rho$ parameter changes notably starting from the value of the aperture radius of 3.96″, so in further calculations we calculated the $Af\rho$ parameter using an aperture radius of 3.96″.

For the SW1 comet (observations of 2006), we calculated $Af\rho = 7325$ cm (for the distances $r = 5.87$ AU and $\Delta = 4.94$ AU). We obtained a high level of dust productivity, which may be explained by the high activity of the comet. Using the results of the observations of December 2007, we obtained a lower value of the dust productivity level $Af\rho = 4637$ cm (for $r = 5.96$ AU and $\Delta = 5.19$ AU). The variation of the $Af\rho$ parameter may be connected with the decrease of the jet number of the comet (in 2006 5–6 jets; in 2007, 3 jets). In a paper by Szabo et al., 2002, for the SW1 comet the authors give a high, relative to our calculations, value of $Af\rho = 16600$ cm for similar distances ($r = 5.9$ AU and $\Delta = 5.1$ AU; observations of 2001), explaining this fact by the outburst activity of the comet.

For the VQ94 comet (observations of 2006, $r = 7.33$ AU and $\Delta = 6.68$ AU), we obtained a value of $Af\rho = 77$ cm. This value is low compared to the SW1 comet, but close to the results given for a distant comet C/NEAT (2001 T4) (from $\log(Af\rho) = 2.5–2.6$ cm) in a paper by Bauer et al. (2003). The comet C/NEAT (2001 T4) was observed at a heliocentric distance of 8.5 AU. The results of observations of Centaur 174P/Echeclus (Rousselot, 2008) demonstrating comet activity at a distance of 13 AU lead to a value of $Af\rho = 10^4$ cm.

To calculate the dust productivity, such parameters as the radius, velocity, density, and albedo of a dust particle are also important. We supposed that the albedo of a dust particle is $p = 0.04$ and the density is 1 g per cm$^3$.

For the calculations, we took a value of the dust particle velocity based on the data given in studies of active comets at long heliocentric distances. An accepted value of the velocity ($v = 30$ m s$^{-1}$) is a certain compromise between values of the velocity for dust particles 1 μm in size for the comets SW1 and Hale–Bopp (50 m s$^{-1}$) accepted for the modeling in papers of Fulle (1992) and Fulle et al., (1998) and the velocity of a dust particle 10 μm in size (10.7 m s$^{-1}$) for the comet C/1999 (Skiff) in a paper by Korsun and Chörny (2003).

To calculate the dust productivity rate, we took a dust particle radius value of $a = 0.005$ mm. This value was accepted from the following ideas. During the long observational time of the SW1 and VQ94 comets, their significant activity remains but distinct dust tails are not observed. The activity in the SW1 and VQ94 comets is observed as jets. This situation is possible if one assumes that from the surface dirty icy dust particles less than ten microns in size are ejected. This assumption is based on the results of a study by Mukai (1986). In this paper, the author says that the lifetime of dirty icy dust particles with a radius more than 0.01 mm is $10^8$ days, when for similar dust particles of a radius from 0.001 to 0.01 mm this time decreases down to few hours. Hence, one may assume that since SW1 and VQ94 have no dust tails, then their atmospheres contain dirty icy dust particles less than 0.01 mm in size. In the opposite situation, a dust tail might form.

Accepting the aforementioned conditions, we calculated the dust productivity rate for the SW1 and VQ94 comets. For the SW1 comet, the value of $Q_{dust}$ is 365 kg s$^{-1}$ for the observations of 2006 and $Q_{dust}$ = 182 kg s$^{-1}$ for the observations of 2007. For the VQ94 comet, we obtained values of $Q_{dust}$ = 6.6 kg s$^{-1}$, which is comparable with the values for the comet C/NEAT (2001 T4) (Bauer et al., 2003) from $Q_{dust} = 10^{-2}$ kg s$^{-1}$ to $Q_{dust}$ = 20 kg s$^{-1}$. The value of the dust productivity for Centaur 174P/Echeclus is 86 kg s$^{-1}$ (Rousselot, 2008).

COMA MORPHOLOGY

Obtained from spectral observations, the surface profiles of the dust and CO$^+$ for the VQ94 comet have significant differences (Korsun et al., 2008). These results became a reason to obtain images of VQ94 in particular filters to have an opportunity to pick out and analyze an image of the comet in CO$^+$ without the dust shell. For this task, we used medium-band filters SED415 ($\lambda$ = 4203 Å, FWHM = 212 Å) and SED537 ($\lambda$ = 5309 Å, FWHM = 169 Å) available in the SCORPIO device. The SED415 filter allows one to pick out the CO$^+$ emission with an addition of the dust component; the SED537 can pick out only the continuum. Consequently, after the necessary reductions, the images were summed for each filter. An image of the coma was obtained by subtraction of the summed image in the SED537 filter from the summed image in the SED415 filter. The subtraction was performed taking into account the differences in the transparency degree of each filter and differences in the continuum level for different wavelengths. The necessary correction was performed by the assumption that the comet continuum and solar spectrum are similar. An additional correction was obtained on the basis of the reddening gradient value taken from the VQ94 spectrum. The image of the dust coma obtained with the SED537 filter and the picked-out CO$^+$ coma are shown in Fig. 1

In Fig. 1, it is seen that the photometric maxima of the dust and CO$^+$ coma components are shifted relative to each other. The value of the shift is 1.4″ (7.44 × 10$^3$ km) in the direction (denoted by vector $J$ in Fig. 1) which is deflected from the direction to the Sun by 63° when the position angle of the comet prolonged radius vector is 193°. We also can see that the structures of the dust and CO$^+$ are different from each other. It is seen that the morphology of the images in both of the filters is different in orientation, the degree of the contour stretch, and their shape. The observed isophote stretch of the dust component is more extensive than for the ionic component. The direction of the stretch is also different for both components.

For the SW1 comet, we obtained images using the CO$^+$ ($\lambda$ = 4280 Å, FWHM = 40 Å) filter, instead of the medium-band SED415 filter, and the SED537 ($\lambda$ = 5309 Å, FWHM = 169 Å) filter to observe the continuum. We used the aforementioned scheme of observations, data reduction, and image subtraction to obtain the pure CO$^+$ coma of the SW1 comet. Contours of the dust and the CO$^+$ coma of the SW1 comet are shown in Fig. 2.

In Fig. 2, one can see that for the SW1 comet the photometric maxima are not shifted relative to each other, as it was for VQ94. The structures of the dust and CO$^+$ are different in the shape and degree of the stretch. The ionosphere of the comet is stretched in the direction opposite to the comet motion vector.

To investigate the morphological structures, we obtained images of SW1 with the broadband $R$ filter in December 2006 and with the medium-band SED537 filter in November 2007. All of the images underwent all of the necessary reductions. To pick out low-contrast structures in the images, we used several digital filters. To pick out jets, the following digital filters were used: Larson–Sakima (1984), unsharping mask, and Gauss blurring. To exclude spurious features when interpreting the obtained images, each filter mentioned above was used separately for the particular image. This technique was used to pick out structures of the comet C/2002 C1 Ikeya–Zhana in a paper by Manzini et al., (2007) and gave good results.

In Fig. 3a, it is seen that after using the digital filters for the SW1 images obtained in December 2006, at least six jets appeared. The orientation of the jets is different and has no particular direction.

In the images of SW1 obtained in November 2007 (Fig. 3b), after using the digital filters we could pick out three jets. All of the jets are directed approximately toward the Sun.

Comparing two images, we can conclude that the orientation and number of jets varies over the course of time for the SW1 comet.

DISCUSSION

The use of modern automatic telescopes (LINEAR, NEAT) has led to the discovery of a new class of comets which are active beyond Jupiter's orbit. In turn, the use of large ground-based telescopes and modern radiation detectors allowed us to investigate in detail the atmospheres and tails of those comets.

Since 2006, we have carried out regular observations of a series of distant comets with the 6-meter telescope of the SAO RAS in the frame of the program of spectral and photometric investigations of comets demonstrating high activity at long heliocentric distances. For this purpose, we selected comets having perihelion distances longer than 5 AU and demonstrating extensive tails.

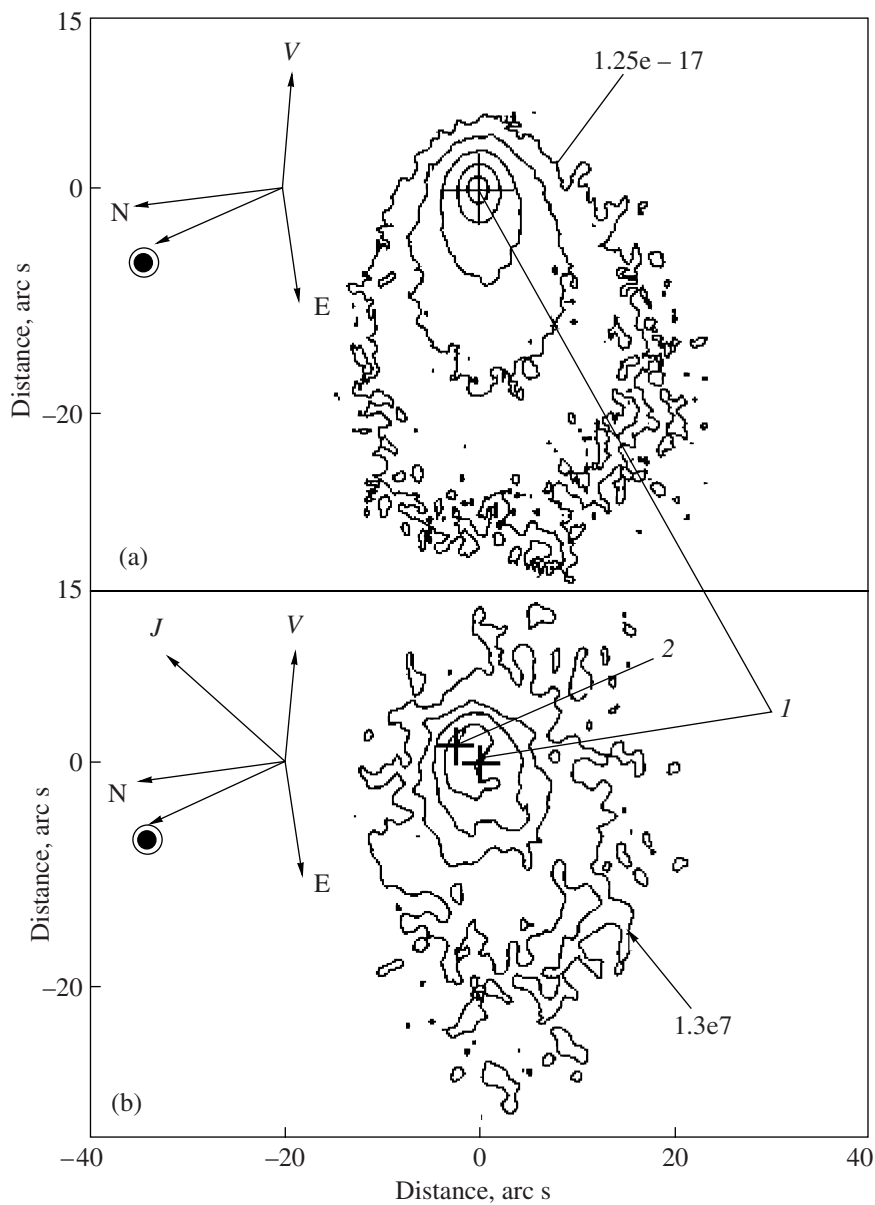

**Fig. 1.** The surface contours of the dust (a) and $CO^+$ (b) coma of the VQ94 comet are shown. Images of the dust coma are given in units of erg/(cm$^2$ s Å), while the images in $CO^+$ are given in units of the number of particles per cm$^2$ using the excitation coefficient for $CO^+$ (Lutz et al., 1993). The direction to the north, east, and shift directions of the photometric centers of the dust and $CO^+$ components (*J*); the direction of the comet motion (*V*); and the direction toward the Sun are shown. Numbers *1* and *2* denote the photometric maxima of the dust and $CO^+$ components, respectively.

During the observation, two comets, C/2002 VQ94 (LINEAR) and 29P/Schwassmann–Wachmann-1, drew our attention due to their activity which is not specific for comets from our sample. Unlike the objects we observed, those two comets demonstrate an asymmetric coma with such features as jets and have no tails.

Both of the comets (C/2002 VQ94 (LINEAR) and 29P/Schwassmann–Wachmann-1) have not only similar morphological features, but a similar chemical composition. If comets from our sample having extensive tails mostly do not demonstrate emissions in their spectra (only the continuum is observed), then SW1 and VQ94 demonstrate bands of $CO^+$, $N_2^+$, and CN in their spectra. The $C_3$ band was identified only in the spectrum of VQ94 (Korsun et al., 2006), but in further observations it was absent (Korsun et al., 2008).

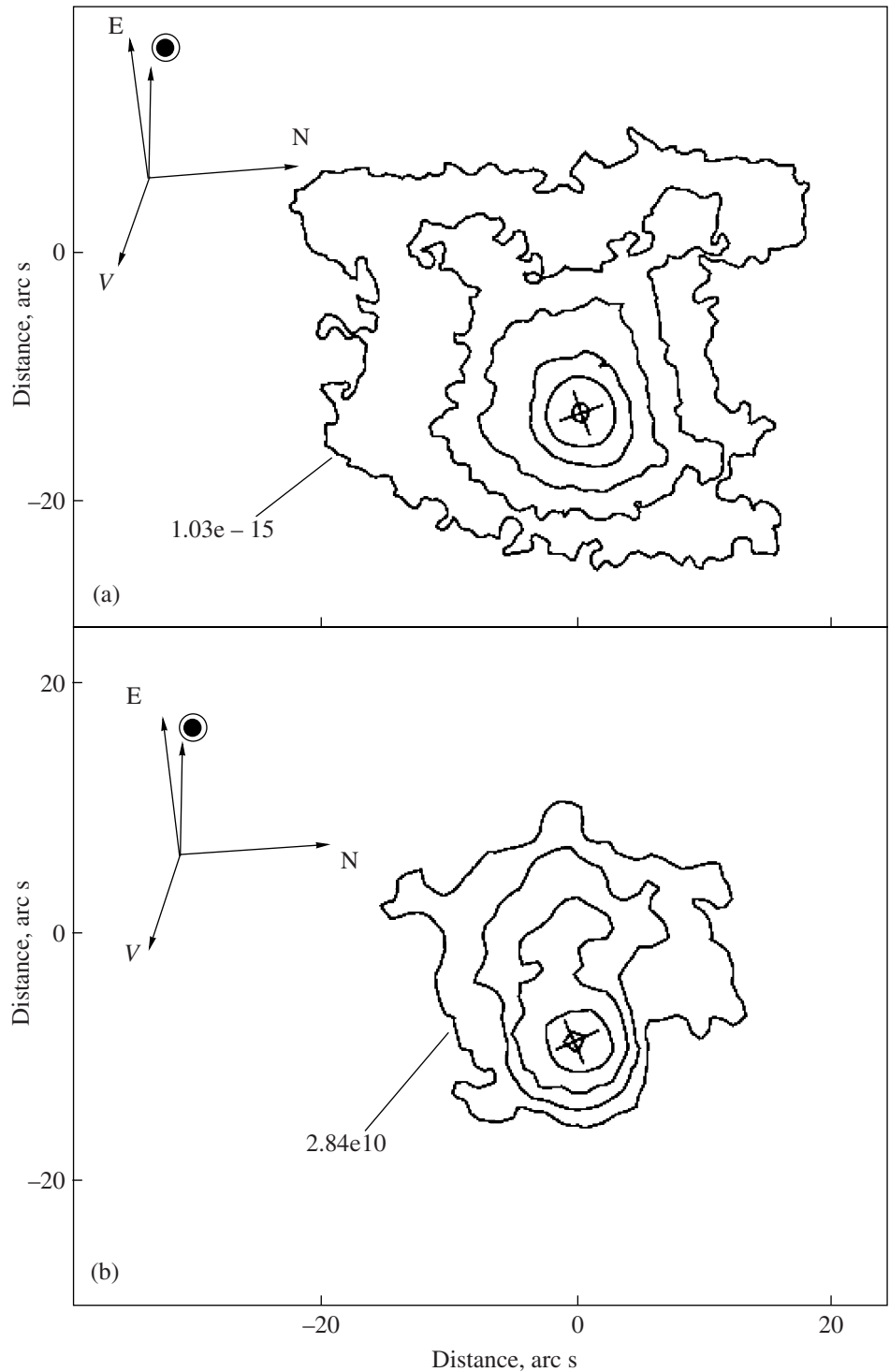

**Fig. 2.** Images of the dust coma are given in units of erg/(cm² s Å), while the images in CO⁺ are given in units of the number of particles per cm² using the excitation coefficient for CO⁺ (Lutz et al., 1993). The direction to the north, east, the direction of the comet motion ($V$), and direction toward the Sun are shown.

Using the results of observations with the broadband filters ($B$, $V$, and $R$), we estimated the color indices for SW1 and VQ94. For the SW1 comet, the color indices were $B-V = 0.64 \pm 0.09$ and $V-R = 0.54 \pm 0.08$. That shows that the comet continuum is redder than the solar continuum. For VQ94, the color index was $V-R = 0.41 \pm 0.07$. These results show that the continuum color of the comet is close to neutral. Spectral observations of these comets (Korsun et al., 2008) show that in the spectral range of 4000–5000 Å, the reddening gradient for the VQ94 comet does not exceed 10%, when for the SW1 comet it is 30%.

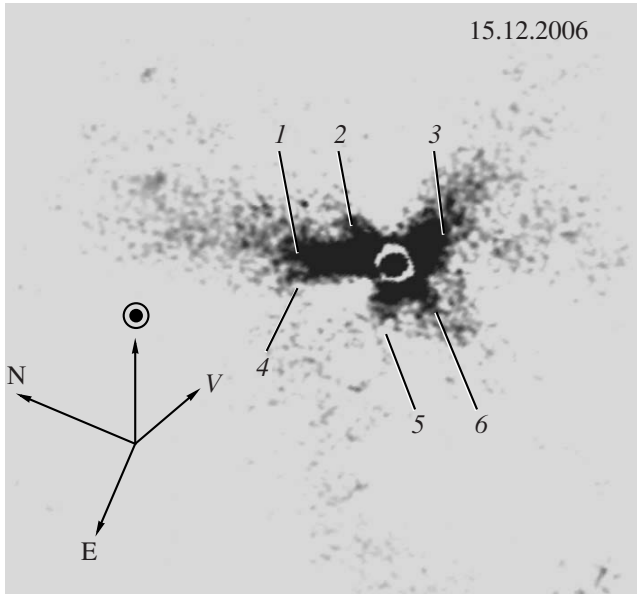

**Fig. 3.** Summed images of the SW1 comet processed using digital filters for two observational runs December 2006

The determination of the dust particle number in the line of sight is a complex task. This is due to the fact that the comet brightness depends on the phase angle, the size distribution of the particles, the albedo of dust particles, etc. The distinguishing of the influence on the comet brightness of the albedo and size of the dust particles is a complex task as well; nonetheless, one may obtain some idea about the measure of the dust productivity from the scattered light flux in the continuum. For our objects, the main problem of the determination of the number of dust particle in the line of sight is a violation of the isotropy of the outflow in the atmospheres of the distant comets, since there the outflow is more concentrated in the jets.

To estimate the amount of dust carried away from the comet surface of SW1 and VQ94 in the frame of some assumptions mentioned above, we calculated the $Af\rho$ parameter for them using images in the continuum.

According to our data, the value of $Af\rho$ for SW1 was 7325 cm at a distance of $r = 8.87$ AU (in 2006) and 4637 cm at a distance of $r = 5.96$ (in 2007). It is seen that the value of $Af\rho$ for SW1 changes over the course of time. The high value of $Af\rho$ can probably be explained by the outburst activity of the comet during its observations.

For the VQ94 comet, we obtained a value of $Af\rho$ equal to 77 cm when it was at a distance of $r = 7.33$ AU. The $Af\rho$ parameters differ significantly for SW1 and VQ94. The obtained values of $Af\rho$ allowed us to estimate the dust productivity rate for SW1 and VQ94. According to our calculations, the dust productivity rate $Q_{dust}$ was 365 kg s$^{-1}$ for the observations of 2006 and $Q_{dust} = 182$ kg s$^{-1}$ for the observations of 2007.

For the VQ94 comet, we obtained a value of $Q_{dust} = 6$ kg s$^{-1}$. This result is comparable with values of the dust productivity rate of comets of the Jupiter family (A'Hearn et al., 1995).

An analysis of the surface profiles of the dust and CO$^+$ obtained from spectral observations for the C/2002 VQ94 (LINEAR) comet (Korsun et al., 2008) showed that they significantly differ from each other. These results became the reason to investigate the morphology of the C/2002 VQ94 (LINEAR) and 29P/Schwassmann–Wachmann-1 comets.

On the basis of the SW1 and VQ94 images obtained with the medium-band filters, we picked out and analyzed the images of the comets in CO$^+$, excluding the dust component. We found that the photometric maxima of the dust and CO$^+$ coma components of VQ94 are shifted relative to each other. The value of the shift is 1.4″ ($7.44 \times 10^3$ km) in the direction which is deflected from the direction to the Sun by 63°.

As for the photometric maxima of the dust and CO$^+$ coma for the SW1 comet, they are not shifted relative to each other. For both comets, the structures of the dust and CO$^+$ are different in the shape and the degree of the stretch of the isophotes. The ionosphere of the comets is stretched in the direction opposite to the comets motion direction. Such a direction of the stretch may be connected with the interaction between the comet and interplanetary medium.

As for the question of the presence of the CO$^+$ and N$_2^+$ lines in the SW1 and VQ94 spectra, it is still unanswered, since at heliocentric distances where the comets were observed, the photo-ionization process of neutral CO runs slowly. And, probably, to explain the ionization of the neutral molecules, one needs to use other or additional mechanisms.

In turn, the presence of a well-developed system of CO$^+$ bands in the spectra of these comets may suppose the presence of a large amount of CO ice in their nuclei, and the sublimation of this ice may be respectable for the formation of comas and tails of these comets at long heliocentric distances. The probable sources of the activity of distant comets and comets at long heliocentric distances have been discussed for a long time. There were attempts to search and find neutral CO molecules in distant comets. Indeed, the CO emissions were found using millimeter range telescopes in the comas of SW1 at a heliocentric distance of ~6 AU (Senay, Jewitt, 1994; Crovisier et al., 1997; Festou et al., 2001), C/1995 O1 Hale–Bopp (Biver et al., 1999; Bockelee-Morvan et al., 2000) at a heliocentric distance of ~8 AU, and the Centaur 2060 Chiron at a heliocentric distance of 8.5 AU (Vumak, Stern, 1999). In addition, for the SW1 comet it was found that the sublimation rate of the CO ice may be enough to explain the observational activity of this comet (Senay, Jewitt, 1994). These observational data and the data we obtained may be evidence that in the Solar System some objects rich in CO and $N_2$ exist. Since according to the laboratory data (Notesco, Bar-Nun, 2005; Notesco et al., 2003; Bar-Nun et al., 2007) objects with a significant abundance of CO and $N_2$ ice might form at a temperature lower than 25 K, it is possible to say that either these objects are formed at significant distances from the Sun or the matter of the protosolar nebula is conserved in them. Hence, a detailed study of such objects is important to understand the scenario of the Solar System's formation.

To investigate the probable structures in the coma of SW1, we used images obtained with the broadband *R* filter. To pick out low-contrast structures in the images of the dust coma of SW1, we used several digital filters: Larson–Sakima (1984), unsharping mask, and Gauss blurring. For comet SW1, we picked out six jets in the filtered image obtained in December 2006 and three jets in the image of November 2007. Filtered images of SW1 obtained in different observational periods showed that the comet activity indeed appears as powerful jets and their number varies in the course of time. The orientation of the jets is different and has no particular direction in the observations of 2006, but in the observations of 2007 all of the jets are directed toward the Sun.

REFERENCES

Vumak, M. and Stern, S.A., Obnaruzhenie gazovoi emissii monookisi ugleroda na 2060 Khirone, *Astron. Vestn*, 1999, vol. 33, no. 3, pp. 216–221.

Kartasheva, T.A. and Chunakova, N.M., Spektral'naya prozrachnost' atmosfery v SAO AN SSSR v 1974-1976 gg, *Astrofiz. Issled. (Izv. SAO)*, 1978, vol. 10, pp. 44–51.

Kiselev, N.N. and Chernova, G.P., Fotometricheskie i polyarizatsionnye nablyudeniya komety Shvassmana-Vakhmana I vo vremya vspyshek, *Pis'Ma V Astron. Zhurn*, 1979, vol. 5, no. 5, pp. 294–299.

*Afanasiev V*, London: Moiseev A.V. The SCORPIO universal focal reducer of the 6-m telescope // Astron. Lett, 2005.

*A'Hearn M.F., Schleicher D.G., Millis R*, London: et al. Comet Bowell, 1980.

*A'Hearn M.F., Millis R*, London: Schleicher D.G., et al. The ensemble properties of comets: Results from narrowband photometry of 85 comets, 1976.

Bauer J.M., Fernandez Y.R., Meech K.J. An Optical survey of the active Centaur C/NEAT (2w001 T4) //Publ. Astron. Soc. Pacific. 2003. V. 115. P. 981–989.

Bar-Nun A., Notesco G., Owen T. Trapping of $N_2$, CO and Ar in amorphous ice // Icarus. 2007. V. 190. P. 655–659.

Biver N., Bockelee-Morvan D., Colom P., et al. Long term evolution of the outgassing of Comet Hale-Bopp from radio observations // Earth, Moon, and Planets. 1999. V. 78. P. 5–11.

Bockelee-Morvan D., Lis D.C., Wink J.E., et al. New molecules found in comet C/1995 O1 (Hale-Bopp). Investigating the link between cometary and interstellar material // Astron. and Astrophys. 2000. V. 353. P. 1101–1114.

Capria M. Sublimation mechanisms of comet nuclei // Earth, Moon, and Planets. 2002. V. 89. № 1. P. 161–178.

Cochran A., Barker E.S., Cochran W. Spectrophotometric observations of P/Schwassmann-Wachmann 1 during outburst // Astron. J. 1980. V. 85. P. 474-477.

Cruikshank D.P., Brown R.H. The nucleus of comet 29P/Schwassmann-Wachmann 1 // Icarus. 1983. V. 56. P. 377.

Crovisier J., Leech K., Bockelee-Morvan D., et al. The infrared spectrum of comet Hale-Bopp // ESA Publ. Div. (ESA SP-419). 1997. P. 137.

*Drilling J.S., Landolt A.U. Allen'S Astrophysical Quantities / Ed Cox A.N*, New York: AIP, 2000.

Festou M.C., Gunnarsson M., Rickman H., et al. The activity of comet 29P/Schwassmann-Wachmann 1 monitored through its CO J = 2 ⟶ 1 radio line // Icarus. 2001. V. 150. Iss. 1. P. 140-150.

Fulle M. Dust from short-period comet P/Schwassmann-Wachmann 1 and replenishment of the interplanetary dust cloud // Nature. 1992. V. 359. № 6390. P. 42–44.

Fulle M., Cremonese G., Bohm C. The preperihelion dust environment of C/1995 O1 Hale-Bopp from 13 to 4 AU // Astron. J. 1998. V. 116. Iss. 3. P. 1470–1477.

Hartmann W.K., Cruikshank D.P., Degewij J. Remote comets and related bodies - VJHK colorimetry and surface materials // Icarus. 1982. V. 52. P. 377-408.

Green D.W.E. Comet C/2002 VQ94 (LINEAR) // IAUC. 2003. 8194.

Gronkowski P., Smela J. The cometary outbursts at large heliocentric distances // Astron. and Astrophys. 1998. V. 338. P. 761-766.

Jewitt D. The persistent coma of comet 29P/Schwassmann-Wachmann 1 // Astrophys. J. 1990. V. 351. P. 277–286.

Jewitt D. Kuiper belt objects // Annu. Rev. Earth and Planet. Sci. 1999. V. 27. P. 287-312.

Jewitt D. A first look at the Damocloids // Astron. J. 2005. V. 129. P. 530.

Jockers K., Bonev T., Ivanova V., Rauer H. First images of a possible $CO^+$-tail of Comet P/Schwassmann-Wachmann 1 observed against the dust coma background // Astron. and Astrophys. 1992. V. 260. ‹ 1-2. P. 455-464.

Korsun P.P., Chorny G.F. Dust tail of the distant comet C/1999 J2 (Skiff) // Astron. and Astrophys. 2003. V. 410. P. 1029–1037.

Korsun P.P., Ivanova O.V., Afanasiev V, London: Cometary activity of distant object C/, 2002.

Korsun P.P., Ivanova O.V., Afanasiev V, London: C/, 2002.

Larson S., Sekanina Z. Coma morphology and dust-emission pattern of periodic Comet Halley. I - High-resolution images taken at Mount Wilson in 1910 // Astron. J. 1984. V. 89. P. 571–578.

Luu J., Jewitt D. Color diversity among the Centaurs and Kuiper belt objects// Astron. J. 1996. V. 112. P. 2310.

Lutz B, London: Womack M., Wagner R.M. Ion abundances and implications for photochemistry in comets Halley, 1986.

Manzini F., Schwarz G., Cosmovici C.B., et al. Comet Ikeya-Zhang (C/2002 C1): Determination of the rotation period from observations of morphological structures // Earth, Moon, and Planets. 2007. V. 100. № 1–2. P. 1–16.

Marsden B.G. MPEC 2002-V71. 2002.

Meech, K.J., Belton, M.J.S., Mueller, B.E.A., et al., Nucleus Properties of P/Schwassmann-Wachmann 1, *Astron. J.*, 1993, vol. 106, pp. 1222 –1236.

Meech K.J., Weaver H.A. Unusual comets as observed from the Hubble Space Telescope // Earth, Moon, and Planets. 1995. V. 72. Iss. 1-3. P. 119–131.

Mukai T. Analysis of a dirty water-ice model for cometary dust // Astron. and Astrophys. 1986. V. 164. № 2. P. 397–407.

Neckel H., Labs D. The solar radiation between 3300 and 12500? // Sol. Phys. 1984. V. 90. P. 205-258.

Notesco G., Bar-Nun A., Owen T. Gas trapping in water ice at very low deposition rates and implications for comets // Icarus. 2003. V. 162. P. 183-189.

Notesco G., Bar-Nun A.A ≤25 K temperature of formation for the submicron ice grains which formed comets // Icarus. 2005. V. 175. P. 546–550.

Oke J.B. Faint spectrophotometric standard stars // Astron. J. 1990. V. 99. P. 1621–1631.

Parker J.W. Distant ECOs // The Kuiper Belt Electronic Newsletter. 2003. № 32.

Prialnik D., Bar-Nun A. Crystallization of amorphous ice as the cause of Comet P/Halley's outburst at 14 AU // Astron. and Astrophys. 1992. V. 258. 2. P. L9-L12.

Prialnik D. Modeling the comet nucleus interior // Earth, Moon, and Planets. 2002. V. 89. ‹ 1. P. 27-52.

Rauer H., Arpigny C., Boehnhardt H., et al. Optical observations of comet Hale-Bopp (C/1995 O1) at large heliocentric distances before perihelion // Science. 1997. V. 275. № 5308. P. 1909-1912.

Roemer E. An outburst of comet Schwassmann-Wachmann 1 // Publ. Astron. Soc. Pacific. 1958. 70. P. 272.

Rousselot P. 174P/Echeclus: A strange case of outburst // Astron. and Astrophys. 2008. V. 480. Iss. 2. P. 543–550.

Sekanina Z., Larson S.M., Hainaut O., et al. Major outburst of periodic Comet Halley at a heliocentric distance of 14 AU // Astron. and Astrophys. 1992. V. 263. № 1–2. P. 367.

Senay M.C., Jewitt D. Coma formation driven by carbon-monoxide release from Comet Schwassmann-Wachmann:1 // Nature. 1994. V. 371. № 6494. P. 229.

Stansberry J.A., Van Cleve J., Reach W.T., et al. Spitzer observations of the dust coma and nucleus of 29P/Schwassmann-Wachmann 1 // Astrop. J. Suppl. Ser. 2004. V. 154. P. 463–468.

Szabó Gy.M., Kiss, L.L: S'arneczky K., Szil'adi K. Spectrophotometry and structural analysis of 5 comets // Astron. and Astrophys, 2002.

Tegler S.C. Romanishin W., Consolmagno G.J. Color patterns in the Kuiper belt: A possible primordial origin // Astrophys. J. 2003. V. 599. P. L49–L52.

Trigo-Rodriguez J.M., Garcia-Melendo E., Davidsson B.J.R., et al. Outburst activity in comets. I. Continuous monitoring of comet 29P/Schwassmann-Wachmann 1 // Astron. and Astrophys. 2008. V. 485. P. 599–606.

Whipple F, London: Rotation and outbursts of comet 29P/Schwassmann-Wachmann 1 // Astron. J, 1980.